\documentclass[twocolumn]{revtex4}
\usepackage{graphicx,amsmath,amssymb,mathrsfs}
\usepackage{epsfig}

\begin{document}

\title{Axion mass estimates from resonant Josephson junctions}

\author{Christian Beck}

\affiliation{Queen Mary University of London, School of Mathematical Sciences, Mile End Road, London E1 4NS, UK}

\begin{abstract}
Recently it has been proposed that dark matter axions from the galactic halo can produce a small
Shapiro step-like signal in Josephson junctions whose Josephson frequency resonates with the
axion mass [C. Beck, PRL 111, 231801 (2013)].
Here we
show that the axion field equations in a voltage-driven Josephson junction environment
allow for a nontrivial solution where the axion-induced electrical current manifests
itself as an oscillating supercurrent.
The linear change of phase associated with
this nontrivial solution implies the formal existence of a large
magnetic field in a tiny surface area of the weak link region of the junction which makes
incoming axions decay into microwave photons. We derive a condition for the design of Josephson junction
experiments
so that they can act as optimum axion detectors. Four independent recent experiments are discussed in this context.
The observed Shapiro step anomalies of all four experiments consistently point towards
an axion mass of $(110\pm 2) \; \mu$eV.
This mass value is compatible with the recent BICEP2 results
and implies that
Peccei-Quinn symmetry breaking was taking place after inflation.
\end{abstract}
%


\maketitle

\section{Introduction}

About 95\% of the energy contents of the universe appears to be of unknown origin, in the form of
dark matter and dark energy. While there is a lot of astrophysical evidence for the
existence of dark matter and dark energy, a deeper understanding of the physical nature
of these main ingredients of the universe is still lacking. Clearly it is important to
design new experiments on earth that could have the potential to unravel some of the
unknown physics underlying dark matter and dark energy.

At the particle physics level, there are
two main candidates what dark matter could be.
These are WIMPS (weakly interacting
massive particles) \cite{bertone}
and axions \cite{duffy, peccei, wilczek-old, sikivie-old}.
WIMPS are motivated by supersymmetry, whereas axions
are motivated by the solution of the strong CP problem
in QCD.
Various experimental searches to detect WIMPS \cite{lab1,bernabei}
and axion-like particles \cite{lab2,lab3,lsw1,admx1} on the earth
are currently going on.


Very recently, there have been a couple of new suggestions how one could
possibly detect dark matter axions in laboratory experiments on the earth
\cite{graham,prl2013,sikivie2014}.
All these proposals have in common that they are based on relatively small
devices and  that they suggest to look for
small oscillating electric currents induced by axion flow, with a frequency given by the
axion mass.
Proposal 1 \cite{graham} is based on a technique similar to nuclear magnetic
resonance (NMRI), known from medical imaging.
Proposal 2 \cite{prl2013} is based on resonance effects in Josephson junctions.
Proposal 3 \cite{sikivie2014} suggests to use LC circuits cooled down to mK temperatures.
Further interesting proposals are based on
topological magnetic insulators \cite{zhang} and atomic systems \cite{roberts}.

In this paper we present a detailed calculation describing the physics of proposal 2,
starting from the field equations of axion electrodynamics in a Josephson environment.
In contrast to axions in vacuum,
in a Josephson junction the axion has the possibility
to induce electric supercurrents, rather than just ordinary currents.
Our main result presented in this paper is that, besides the trivial solution where the axion passes
through the Josephson junction without interaction, there is a nontrivial
solution to the axion field equations
due to these supercurrents.
We show that the nontrivial solution implies the existence of a huge (formal) axion-flow generated
magnetic field
in a tiny surface area of the weak-link region of the junction, which makes incoming axions decay
into microwave photons.
The axion flow from the galactic halo through the junction then leads to
a small measurable excess current of Cooper pairs,
for which we will derive a concrete formula.
The experimental consequence of this are Shapiro steps \cite{shapiro,tinkham}
generated by axion flow, which are small but observable provided
certain conditions on the design of the Josephson junction are satisfied.
We will derive these conditions explicitly.

An experiment by Hoffmann et al. based on S/N/S Josephson junctions \cite{hoffmann}, discussed in detail in \cite{prl2013},
provided evidence for an axion mass of 110 $\mu eV$ and an axionic dark matter density of about
0.05 GeV/$cm^3$ if interpreted in this way.
Here we will discuss the results of four
different experiments \cite{hoffmann,golikova,he,bae}.
In all four cases small
Shapiro step-like anomalies have been observed that, if interpreted within our theory,
point towards an axion mass of $m_ac^2=(110\pm 2) \mu$eV.

The predicted axion mass value has profound cosmological implications.
If this value
is confirmed by further experiments, it means that the Peccei-Quinn symmetry breaking took
place {\em after} inflation \cite{visinelli2}. Employing the recent results of \cite{visinelli2,bicep}
our result implies that the fractional contribution $\alpha^{dec}$ to the cosmic
axion density from decays of axionic strings and walls is $\alpha^{dec}=0.66 \pm 0.05$.

This paper is organized as follows: In section 2 we write down the axion field
equations in a Josephson junction. The nontrivial solution, where the axion-induced electric
current manifests itself as a supercurrent within the junction, is discussed in section 3. The physical interpretation
of this solution is further worked out in section 4. In section 5 we present a short
calculation how S/N/S Josephson junctions should be designed in order
to serve as optimum axion detectors. Section 6 discusses some experimental candidate signals
seen in various Josephson experiments that could possibly be associated with the
nontrivial solution of section 3.
Section 7 compares our mass estimate from Josephson resonances with
cosmological and astrophysical bounds on the axion mass. Finally, our concluding
remarks are given in section 8.
\


\section{Axion field equations in a Josephson junction}

Let us consider the classical field equations of axion electrodynamics \cite{sikivie-old, sikivie2007, wilczek-old, visinelli}
in a Josephson
junction (JJ) \cite{josephson, shapiro, prl2013, tinkham}. $\theta=a/f_a$ denotes the misalignement angle of the axion field $a$, $\delta$
the electromagnetic phase difference in the JJ. In SI units one has

\begin{eqnarray}
\ddot{\theta} + \Gamma \dot{\theta} -c^2 \nabla^2 \theta +\frac{m_a^2c^4}{\hbar^2} \sin \theta
&=& - \frac{g_\gamma}{4 \pi^2} \frac{1}{f_a^2} c^3 e^2 \vec{E} \vec{B} \label{1} \\
\nabla \times \vec{B} - \frac{1}{c^2} \frac{\partial \vec{E}}{\partial t} = \mu_0 \vec{j} &+&\!\!\!\! \frac{g_\gamma \alpha}{\pi c}
\left(\vec{E} \times \!\!\nabla \! \theta -  \vec{B} \dot{\theta}\right)  \label{2} \\
\nabla \vec{E} &=& \frac{\rho}{\epsilon_0} + \frac{g_\gamma}{\pi} \alpha c \vec{B} \nabla \theta
\\
\ddot {\delta} + \frac{1}{RC} \dot{\delta} + \frac{2eI_c}{\hbar C} \sin \delta &=& \frac{2e}{\hbar C} (I+I_a). \label{4}
\end{eqnarray}
Here
$m_a$ denotes the axion mass, $f_a$ is the axion coupling constant,
$\Gamma$ is a tiny damping constant,
$\vec{E}$ is the electric field, $\vec{B}$ is the magnetic field, $g_\gamma$ is a coupling constant of order 1
($g_\gamma =-0.97$ for KSVZ axions \cite{ksvz1,ksvz2}, $g_\gamma=0.36$ for DFSZ axions \cite{dfsz1,dfsz2}), $\alpha$ is the fine structure constant,
$I_c$ is the critical current of the junction, $I$ an external driving current, $I_a$ is a small axion-induced
electric current in the junction, $R$ is the normal resistance of the junction, and $C$ its capacity.
As usual, $\vec{j}$ and $\rho$ denote electric current and charge densities.

The expected mass of the QCD dark matter axion is in the region $\mu$eV to $m$eV
due to astrophysical and cosmological constraints; the corresponding Compton wave length
is much larger than the typical size of a JJ. Thus we may neglect
spatial gradient terms $\nabla \theta$ in the above equations and consider the axion field as being approximately
spatially homogeneous.
The most important axion contribution for detection
purposes comes from the last term in eq.~(\ref{2}): In a magnetic
field $\vec{B}$, temporal changes $\dot{\theta}$ of the axion angle imply
an axion-induced electric current density $\vec{j}_a$ given by
\begin{equation}
\mu_0 \vec{j}_a = -\frac{g_\gamma \alpha}{\pi c} \vec{B} \dot{\theta}. \label {5}
\end{equation}
Note that this current is in the direction of the magnetic field $\vec{B}$, and not orthogonal
to it, as in ordinary electrodynamics.

Dark matter axions correspond to an oscillating solution of
eq.~(\ref{1}) with $\vec{E} \vec{B}=0$ and $\Gamma$ negligible, given by
\begin{equation}
\theta (t)= \theta_0 \cos (\omega_a t+ const). \label{6}
\end{equation}
The frequency $\omega_a= 2\pi \nu_a =m_ac^2/ \hbar$ is given by the axion mass.
The dark matter energy density due to axions,
$\rho_a$, is related to the amplitude $\theta_0$ of the oscillations by
\begin{equation}
\rho_a =\frac{1}{\hbar^3 c^3} \frac{1}{2} m_a^2 c^4 f_a^2 \theta_0^2.
\end{equation}
This
can be used to eliminate $\theta_0$
as
\begin{equation}
\theta_0= \frac{\sqrt{2c^3 \hbar^3 \rho_a}}{f_a m_ac^2}.        \label{17}
\end{equation}
$\theta_0$ is very small: The astrophysical estimates of dark matter density in the halo \cite{boer}
give something of the order $\theta_0 \sim 10^{-19}$.


Now consider as a suitable axion detector a driven JJ in the voltage stage
which
contains a constant magnetic field $\vec{B}$
near the surface of the weak link (WL) region that points in the
direction of the bias current $I$ of the junction.
At the moment we do not discuss the origin of this magnetic field, it can be an external
magnetic field, though
later we will see that as a consequence
of the field equations axions can formally self-induce such a
magnetic field at the surface of WL.
We denote the distance between the two superconducting electrodes of the junction by $d$, the width of the
superconductors by $w$ and their height by $L$, so that the volume of the weak link
region is $dwL$. The volume of the region where the magnetic field is present is $dwL_1$,
we assume $L_1 << L$, i.e. the magnetic field is only present near the surface of WL.
Let us consider axions from the galactic halo that
enter WL transversally with velocity $v_a$ through the plane spanned up by $w$ and $d$.
The dark matter oscillations (\ref{6})
yield via eq.~(\ref{5}) an axion-induced current $\vec{I}_a$ which couples into
the JJ via eq.~(\ref{4}). $\vec{I}_a$
is given by
\begin{eqnarray}
\vec{I}_a (t)= \vec{j}_a (t) A&=& -\frac{Ag_\gamma \alpha}{\mu_0 \pi c} \vec{B} \dot {\theta} \\
&=&\frac{Ag_\gamma \alpha \omega_a \theta_0}{\mu_0 \pi c} \vec{B} \sin (\omega_a t+const). \label{8}
\end{eqnarray}
Here $A=L_1w$ is the small surface area through which the magnetic field penetrates.
$\vec{I}_a(t)$ is an oscillating current produced by entering dark matter axions coming from outside WL.
What happens inside WL will be discussed in the next section.

\section{A nontrivial solution inside the weak link}

The main difference between axions in vacuum and axions in a JJ is that in a JJ
there is the possibility that electric currents are supercurrents, i.e. they manifest
themselves in form of Cooper pairs that tunnel the weak-link region. This possibility
is also open to electric currents induced by dark matter axions in a magnetic field.
The main assumption in the following is that
the oscillating current (\ref{8}) manifests itself as a supercurrent in the JJ. This means we can write
for $I_a(t)=|\vec{I}_a(t)|$
\begin{equation}
I_a (t) = I_c^a \sin \theta (t), \label{9}
\end{equation}
where $I_c^a$ can be regarded as an axion-generated critical current in the junction,
and the phase $\theta (t)$ grows linearly inside WL. 
Note that $\theta=\theta (\vec{x},t)$ has different behavior for $\vec{x} \in$ WL and $\vec{x} \notin$ WL.
We will later see what the physical
interpretation of this is (in fact inside WL the axion will decay immediately so that
the phase $\theta$ will get the meaning of an ordinary electromagnetic phase). At the moment we just
do the maths, i.e. regard $\theta$ as a solution of the equations and come to the physical interpretation later.

Equality between eq.~(\ref{8}) and (\ref{9})
implies that inside WL the variable $\theta$ evolves in a different way
than outside WL, namely, we have inside WL
\begin{equation}
\theta (t)= \omega_a t+const \label{here}
\end{equation}
and
\begin{equation}
I_c^a=\frac{Ag_\gamma \alpha B \omega_a \theta_0}{\mu_0 \pi c}. \label{11}
\end{equation}
Apparently, coincidence of the
electric currents (\ref{9}), (\ref{8}) produced in a small vicinity inside and outside WL
requires that
the angle variable $\theta (t)$
switches from harmonic oscillations to
linearly increasing behavior modulo $2 \pi$ when entering WL,
still with the same frequency $\omega_a$.
Note that the phase difference $\delta$ of the JJ anyway increases linearly in time, i.e. $\dot{\delta}=\omega_J$,
where $\omega_J=2eV/\hbar$ is the Josephson frequency,
since the JJ is in a driven voltage stage \cite{tinkham}. This means that if $\omega_J=\omega_a$
the linear change of $\delta$ agrees with that of $\theta$, which is
physically absolutely reasonable.

The linear change $\dot{\theta} = \omega_a$ from eq.~(\ref{here})
still has to respect the axion field equations. It couples back into the system via
eq.~(\ref{1}).
Suppose the effective damping constant $\Gamma$ in eq.~(\ref{1}) (at the surface of the junction) satisfies
\begin{equation}
\Gamma >> \omega_a \theta_0. \label{cond}
\end{equation}
This is not a very restrictive
condition, since for axions with a mass of $O(100\mu eV)$ one has the very small value $\omega_a \theta_0 \sim 10^{-8}s^{-1}$.
If this condition is satisfied
then
the first, third and fourth term on the left-hand side of eq.~(\ref{1}) can be neglected,
and eq.~(\ref{1}) reduces to
\begin{equation}
\Gamma \dot{\theta} =- \frac{g_\gamma}{4 \pi^2} \frac{1}{f_a^2} c^3 e^2 \vec{E} \vec{B}. \label{there}
\end{equation}
(If condition (\ref{cond}) is not satisfied then there will be an additional
oscillating component to the $B$-field, otherwise the argumentation is very similar.)
The linear increase $\dot{\theta}=\omega_a\not= 0$ inside WL due to eq.~(\ref{here}) thus formally induces an
axion-generated magnetic field $\vec{B}$ via eq.~(\ref{there}).
Using $E=V/d$, where
$V$ is the external voltage applied to the JJ, as well as using the resonance condition
$m_ac^2=2eV$ between axion mass and Josephson frequency $\omega_J=2eV/\hbar=\dot{\delta}$ (which can be achieved by suitably
choosing $V$)
we get from eq.~(\ref{there})
\begin{equation}
B= -\frac{8 \pi^2 f_a^2 \Gamma d}{g_\gamma c^3 e \hbar}. \label{13}
\end{equation}
We have thus found a nontrivial solution of the field equations
(\ref{1})--(\ref{4}) where there is an axion-generated magnetic field (in the direction
of the bias current) given by (\ref{13})
due to the existence of supercurrents.
Depending on what is assumed for $\Gamma$, this can be a huge magnetic field
(see \cite{prl2013} for some numerical examples), but it is penetrating only through a tiny
surface area $A=L_1w$ so that the flux is reasonably sized.

To properly describe axion electrodynamics in a JJ, we
also need to take into account the probability of axion decay in a strong
magnetic field.
From the Primakov effect one has for
the decay probability of axions in a magnetic field of strength $B$ \cite{sikivie2007}
\begin{equation}
P_{a\to \gamma}= \frac{1}{16\beta_a}(g_\gamma  \; Bec \;L)^2 \frac{1}{\pi^3f_a^2} \alpha
\left( \frac{\sin \frac{qL}{2\hbar}}{\frac{qL}{2\hbar}} \right)^2 .
\end{equation}
Here $L$ is the length of the detector, $q$ is the axion-photon momentum transfer,
and $\beta_a=v_a/c$.
In particular, for the
length scale $L_1$ within which the axion decays with probability $P_{a \to \gamma}=1$
one has (for $qL_1 <<\hbar$)
\begin{equation}
L_1=\frac{\hbar c^2}{f_a \Gamma d} \sqrt{\frac{\beta_a}{4\pi \alpha}} .\label{15}
\end{equation}

Clearly, the axion-generated $\vec{B}$ field is only present in the area where
the axion still exists and has not yet decayed. This suggests to use for the area
$A$ in equation (\ref{11}) the value $A=L_1w$, where $L_1$ is given by (\ref{15}).
Putting eq.~(\ref{13}), (\ref{15}) and (\ref{17})
into eq.~(\ref{11}) a remarkable simplification takes place and one finally ends up with the simple formula
\begin{equation}
I_c^a=\sqrt{\frac{\rho_a v_a}{h \alpha}}w \cdot 2e. \label{ic1}
\end{equation}
Note that physically unmeasurable quantities like the formal
huge magnetic field concentrated in a tiny surface area as well as the unknown constant
$\Gamma$ have all dropped out,
and the critical current $I_c^a$ is basically determined by the dark matter axion velocity $v_a$
relative to the junction and the axionic dark matter density near the earth $\rho_a$, as well as the fundamental constants
 $h=2\pi \hbar$, $\alpha$ and $2e$ (the charge of Cooper pairs).

\section{Physical interpretation of the nontrivial solution}

A linearly increasing phase in ordinary JJ physics means tunneling Cooper pairs and emission
of Josephson radiation \cite{tinkham}. As we have seen, if the axion is still assumed to be present
in WL, then a linearly increasing phase $\dot{\theta}=\omega_a=\omega_J$ implies the existence of
a huge magnetic field given by eq.~(\ref{13}). Alone for energetic reasons, such a huge field
cannot be present in the junction. The only sensible conclusion is that the axion immediately
decays when entering the junction, it tunnels the junction \cite{prl2013}.
If the axion is still present in WL, then the huge magnetic field induced by (\ref{13}) will make it decay immediately.
If the axion is not present in WL, then there is no huge magnetic field, and hence no energetic problem with that field.
During the tunneling process the phase $\theta (t)$ just
gets the meaning of an ordinary electromagnetic phase in the junction, which obeys standard
type of SQUID physics \cite{prl2013}. The tunneling axion still produces a measurable effect
due to an additional contribution to
the critical current given by (\ref{ic1}).

In calculating this critical current, all singular (unmeasurable) quantities
such as the huge formal magnetic field have dropped out. 
We used the formal magnetic field mainly as a mathematical tool
to calculate the axion-generated effects in WL from the field equations.
In experiments one should look for
a small peak-like structure of the differential conductance
at a particular voltage $V$, produced by the increase in the critical current.
This voltage satisfies
$2eV=m_ac^2=\hbar \omega_a$.
Putting into eq.~(\ref{ic1})  typical numbers for the estimated density $\rho_a$ and velocity $v_a$ of dark matter axions
in the galactic halo relative to the Earth one sees that the predicted contribution of axions to
the critical current is small but perfectly measurable, of the order of $\sim10^{-8}$A.
Our physical interpretation of the linearly
increasing phase $\theta (t)$ is that the axions
passing through WL decay via Josephson radiation, triggering at the same time
additional Cooper pairs to flow. These additional Cooper pairs yield a small but measurable effect
if $\omega_J=\omega_a$.


 In a recent preprint \cite{wilczek2014} it was proposed that axions can have different interactions with
 Cooper pairs in a superconductor than in the vacuum case, possibly stronger ones, since electron number is not conserved.
 This is in line with the ideas presented here: In our consideration
 presented here and in \cite{prl2013}, incoming axions decay
 into microwave photons and trigger at the same time the process of
 Cooper pairs forming out of ordinary
 electrons, thus producing
 additional supercurrents in the voltage-biased junction. The (maximum)
 additional critical current $I_c^a$ as allowed by axion electrodynamics is given by (\ref{ic1}). This critical current $I_c^a$
 is derived from the axion field equation (\ref{2}) and is a current
 of ordinary Cooper pairs through an ordinary
 electromagnetic weak link; it is not related to the existence or non-existence
 of a possible weak link in Peccei-Quinn symmetry as discussed in \cite{wilczek2014}.

 In topological superconductors \cite{witten}, superconducting phases couple to electromagnetic
 fields with an axionic topological coupling, and chiral vortex lines can be Josephson junctions.
 The corresponding field theory was developed in \cite{witten}. Axions can also induce nonlinear modulations
 of the electromagnetic field in topological insulators \cite{zhang}. Generally
 it is to be expected that axionic dark matter is much more likely to leave a detectable signal in complex
 systems of condensed matter physics than in vacuum.

\section{The optimum axion detector}



Coming back to the simple case of axions passing through WL of a JJ,
 as with any critical current, $I_c^a$ represents the {\em maximum} electric supercurrent one can expect to see
 due to axionic dark matter passing through the junction, an idealized situation, as
 allowed by the classical field equations. In practice, the
 current will often be smaller, depending on the experimental situation and the
 geometry of the Josephson junction used.
 To work this out further, let us denote by $A^*$ the surface area of WL perpendicular to axion flow, which
 for orthogonally entering axions is
 given by $A^*=wd$.
 So far we did mainly
 do classical axion electrodynamics. But one also needs to take
 into account the particle nature of axions and Cooper pairs. Following the ideas of \cite{prl2013},
 we may assume that each axion entering WL triggers the flow of $N$ Cooper pairs (Fig.~2 in \cite{prl2013}).
 In an S/N/S junction $N$ is related to the number of Andreev reflections and given by
  \begin{equation}
 N \approx \frac{2 \Delta}{eV} +1 \label{N}
 \end{equation}
 where $\Delta$ is the gap energy of the superconductor \cite{hoffmann,prl2013}.
 The maximum observable supercurrent is constrained by the geometry of the JJ
 and given
 by the number of axions hitting the
 WL region of surface area $A^*$ per time unit, multiplied by $2eN$:
 \begin{equation}
 \hat{I}_c^a =\frac{\rho_a}{m_ac^2} v_a A^* \cdot 2eN \label{ic2}
 \end{equation}
 For an optimum S/N/S axion detector,
 both formulas (\ref{ic1}) and (\ref{ic2}) should be valid, i.e. $I_c^a \approx \hat{I}_c^a$.
 By equating them we obtain
  \begin{equation}
 d N \approx \frac{m_ac^2}{\sqrt{\alpha h \rho_a v_a}}. \label{dn}
 \end{equation}
 The right-hand side is just determined by astrophysical dark matter properties, whereas the
 left-hand side yields a relation for the detecting JJ experiment. For the Aluminium junction used
 by Hoffmann et al. \cite{hoffmann}, one can readily check that
 condition (\ref{dn}) is satisfied. This experiment thus provides an optimum axion detector.
 If one wants to use other S/N/S Josephson junction, say with a higher gap
 energy $\Delta$, then naturally $d$ must be chosen smaller.
For the dark matter parameters advocated in \cite{prl2013}, one obtains a characteristic length scale of
 $dN \approx 6 \mu m$.

 \section{Experimental candidate signals in Josephson junction experiments}

 Let us now discuss the experimental consequences of our theory. As previously discussed in \cite{prl2013},
 the decaying axions produce photons and these produce small axion-induced Shapiro steps,
 which are measurable for junctions with sufficiently large $A^*$.
 The main Shapiro step
 occurs at a voltage given by $V_a=m_ac^2/2e$, other integer multiples of $V_a$ may
 also occur if the axion-induced Cooper pair flow intensity is high. The typical step size
 should be given by eq.~(\ref{ic2}).

 First, let us discuss Hoffmann et al.'s experiment \cite{hoffmann}, based on
 Al-Cu-Al S/N/S junctions.
  Their measurement of a Shapiro step-like feature of unknown origin at $V_a=55 \mu$eV  in
\cite{hoffmann} was used in \cite{prl2013} to estimate the axion mass as
$m_ac^2=2eV_a=110 \mu eV$ and the axionic dark matter density near the earth as being $\rho_a=0.051 GeV/m^3$,
assuming orthogonal axion flow.
The velocity $v_a$ of axions traveling through the JJ was assumed to be given by the
value $v_a=2.3 \cdot 10^5 m/s$ (the velocity of the earth relative to the galactic halo).
Let us now look at other experiments as well.

 In the experiment of Golikova et al. \cite{golikova}, based on Al-(Cu/Fe)-Al microbridges,
 a
 double-peak peculiarity of the measured differential resistance is observed
 (Fig.~4 in \cite{golikova}), with one rather constant peak occurring at $(52 \pm 5) \mu$V,
 whereas the other peak position near 75$\mu$V is dependent on the length of the sample and the applied magnetic field.
 A possible interpretation would be to interpret
 the first (universal) peak as coming from axions and the second peak as being
 due to a minigap produced by the proximity effect. Further measurements are needed to check this.

 He et al. \cite{he} use W-Au-W S/N/S junctions and report the observation
 of a large number of fractional Shapiro steps without externally applied microwave radiation.
 Most Shapiro steps occur in the temperature region 2.8-3.2K,
 and for this temperature region the strongest steps occur at $(53\pm 3) \mu$V (see Fig.~2
 in \cite{he}), which is again the axion
 voltage $V_a$.
 Since the Wolfram superconductors used by He et al. have a gap energy
 $\Delta$ that is larger by a factor 5 as compared to the aluminium superconductors used in \cite{hoffmann} and \cite{golikova},
 the number of Andreev reflections $N$ given in
 eq.~(\ref{N}) is larger, and hence a higher intensity of Cooper pair flow is induced by the incoming axions.
 This may be the reason that a
 larger number of observable Shapiro steps
 is excited in this experiment. In addition, the minigap structure created by the proximity effect may create
 further Shapiro steps in this junction.

 Finally let us discuss another experiment performed by Bae et al.\cite{bae},
 which is quite different from the previous ones, in the sense that
 a high-$T_c$ superconductor is used, and that there is also some
 external forcing with microwaves with a given frequency $\nu_1$.
 Bae et al. \cite{bae}
 investigated the occurence of Shapiro steps when irradiating
 a micron-sized sample of BI-2212, a high-$T_c$ crystal, which contains a stack of about $80$
 intrinsic tunnel Josephson junctions. The superconducting layers of this crystal are separated by
 $d=1.2$nm from each other. The sample used by Bae et al. had width $w=5\mu$m,
 hence in total the area of the WL region is given
 by $A^*=80\cdot dw=4.8 \cdot 10^{-13}m^2$. This effective area is large enough to produce measurable
 axion-induced currents via eq.~(\ref{ic2}). From ~(\ref{ic2}) one obtains the
 prediction $\hat{I}_c^a=$
 $16.4 nA$ if $N=1$ and $\rho_a=0.051$ GeV/$cm^3$ is used.

 Bae et al. \cite{bae} irradiated their
 probe by external microwave radiation of frequency $\nu_1=$  5GHz, 13 GHz,
 18 GHz, 23GHz, and 26 GHz, respectively. Note that 26 GHz $=\nu_a$.
 For all values of $\nu_1$ they observed well-pronounced integer Shapiro steps at voltages $V_n$ given by
 $2eV_n=n h \nu_1$, as expected from the RSJ model \cite{tinkham}, with $n$ integer.
 However, two unexplained peculiarities occured (see Fig.~1, data from \cite{bae}):

 \begin{figure}
 \epsfig{width=8cm, height=5cm, file=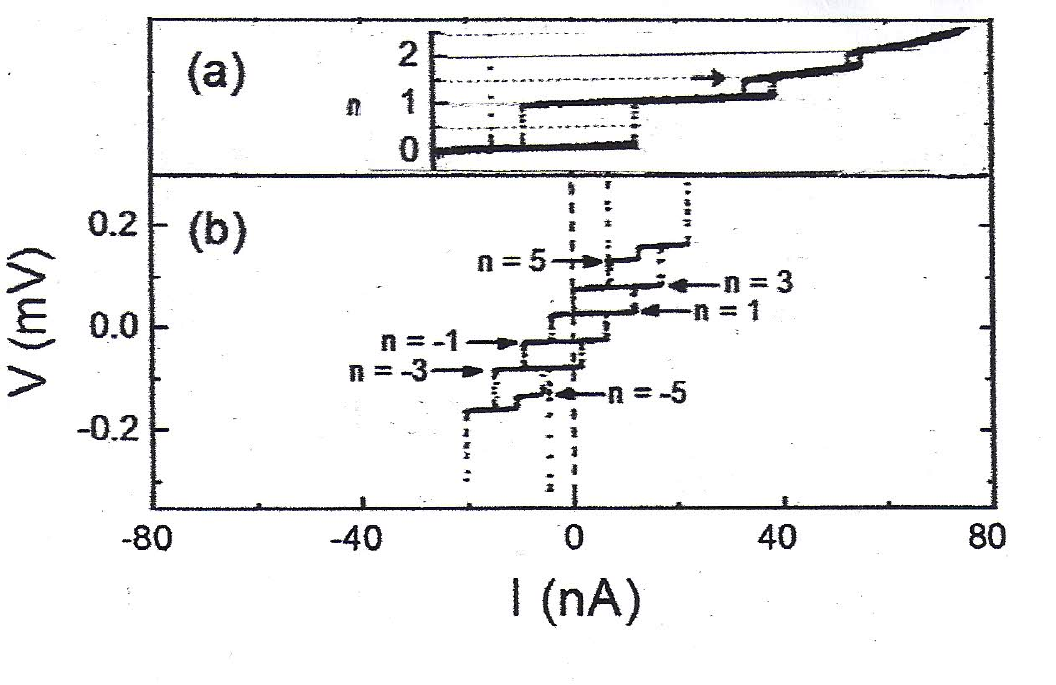}
 \caption{Possible signatures of a 110 $\mu eV$
 axion in the measurements of Bae et al. \cite{bae}: (a) Anomalous Shapiro step (indicated by arrow) occuring at
 the voltage $V_a=55 \mu$V ($n=3/2$) for $\nu_1=18$ GHz and (b) anomalous double-size Shapiro steps
 with voltage differences $55  \mu$V occuring at odd integers for $\nu_1=13$GHz.
 }
  \end{figure}

 1. While for $\nu_1=5,23,26$ GHz only the usual integer Shapiro steps were observed, at the frequency
 $\nu_1=18$ GHz also one additional fractional Shapiro step is seen, formally with $n = 3/2$
 (see Fig.~1a). This unusual
 step occurs at a voltage $(55 \pm 1)\mu$V, and the step size is (for increasing voltage) 16.4 nA.
 While in principle fractional Shapiro steps are possible due to non-sinusoidal contributions
 in the current-phase relation,
 it is very unusual to have only one
 such step (there is none at e.g.\ $n=1/2$ and $n=5/2$, and also none for the other values of $\nu_1$).
 Our physical interpretation is that this unusual step is due to axion flow through the junction
 and that it is stabilized due to the commensurate ratio $3\nu_1=2\nu_a$. The observed step size 16.4nA
 agrees with what is expected from eq.~(\ref{ic2}).

 2. While for $\nu_1=5,23,26$ GHz basically all low-$n$ integer Shapiro steps are observed, for
 $\nu_1=13$ GHz only odd-$n$ Shapiro steps are seen ($n=-5,-3,-1,1,3,5$)
 whereas the even-$n$ Shapiro steps $(n=-4,-2,0,2,4)$ are suppressed,
 so that the voltage difference between neighbored steps is $(55 \pm 1)\mu$V rather
 than the expected 27$\mu$V (see Fig.~1b).
 Due to these double-voltage steps the pattern looks much more similar
 to a pattern generated by a (phase-shifted) frequency of 26 GHz rather than one
 of 13 GHz. Our physical interpretation is that at even $n$ the axion-generated Shapiro steps
 fall onto the ordinary ones generated by the frequency $\nu_1$ and can compensate them, provided they have the same
 magnitude but opposite sign. The observed zero-crossing step sizes
 for $n=-3,-1,1,3$ are indeed 16.4 nA, which would allow for such a destructive interference.

In summary, all four experiments mentioned contain peculiar Josephson resonance effects
associated with the voltage $V_a\approx 55\mu$V, pointing towards
an axion mass of $(110 \pm 2) \mu$eV, where we base our error estimate on the data of \cite{hoffmann} and \cite{bae}.
Additional experimental tests are of course still needed.
A typical
axion Shapiro step is predicted to exhibit small daily and yearly periodic oscillations in intensity, similar as in
searches for WIMPS \cite{bernabei}. The daily oscillations are expected
to come from the fact that the galactic axion
flow relative to the Earth is directed, the Earth rotates but axions produce
the strongest signal if they enter the junction transversally.
We also emphasize that it is
clearly important to extend the search range of other axion search experiments, which are {\em not} based
on JJs, to
the mass region suggested by JJs, $m_ac^2 \sim 0.11 m$eV.
A recent experimental proposal in this direction is \cite{rybka}.

\section{Axion mass estimates}

Let us finally discuss the cosmological consequences of an axion mass value of 0.11meV
that the experiments discussed in the previous section seem to favor.
This value implies that
the Peccei-Quinn symmetry was broken {\em after} the end of inflation \cite{visinelli2,bicep},
at least in the simplest models of cosmological axion production.
Based on the results of \cite{visinelli2} (assuming that axions make up all dark matter),
the mass value $m_ac^2=110\mu$eV translates
into an axion coupling constant of $f_a=5.64 \cdot 10^{10}$GeV, a freeze-out temperature of $T_f=998$ MeV, and
for the fractional contribution $\alpha^{dec}$ to the cosmic axion density from decays of axionic strings and walls
we obtain from the formula
\begin{equation}
\alpha^{dec} = \left( \frac{m_ac^2}{(71 \pm 2 )\mu eV} \right)^{7/6} -1
\end{equation}
derived in  \cite{visinelli2} the prediction $\alpha^{dec}=0.66 \pm 0.05$.

Generally, the rather high energy scale of inflation that seems to be indicated by the recent
BICEP2 results \cite{bicep} puts a lower bound on the QCD axion mass of about 70$\mu$eV \cite{visinelli2,italy}.
If production
of axions from axionic strings is taken into account as well \cite{shellard},
then this lower bound still becomes sharper: Shellard et al. derive in \cite{shellard} for the axion coupling
constant $f_a<7.4 \cdot 10^{10}$ GeV, which is equivalent to $m_a>84\mu$eV.
On the other hand, the axion mass cannot be too
big because otherwise one would violate various astrophysical observational constraints.
For example, the SN1987a supernovae data imply a lower bound on $f_a$ given approximately by $10^{9}$ GeV
\cite{raffelt, turner}, equivalent
to $m_a<6$meV. In \cite{shellard} an even stronger upper bound is derived if
quantum fluctuations during inflation are taken into account, $m_a<1$meV.
A recent analysis of galactic rotation curves \cite{li} based
on an axionic Bose-Einstein condensate as dark matter is consistent with
this range of an axion mass between 0.1 and 1meV, as
well as with estimates of the axionic dark matter density in the halo as
calculated in \cite{prl2013}.

Overall, it appears that the mass value $m_ac^2= 0.11$meV that is suggested by the
Josephson resonances fits very well into the
mass range expected from cosmological and astrophysical considerations after BICEP2.

\section{Conclusion}

In this paper we have presented a detailed derivation
why axions can generate small measurable electric currents in Josephson junctions.
We started from the field equations of axion electrodynamics, plus
the assumption that the axion-induced electric current can manifest itself
as a supercurrent in a Josephson junction.
We found a nontrivial solution of the axion field equations in the weak-link region of the junction,
for which the phase grows linearly in time.
This was interpreted in terms of axions decaying via Josephson radiation and triggering at the
same time additional Cooper pair flow through the junction.
A huge formal magnetic field appeared in our calculations,
making the axions decay if they are still present at the surface of the junction,
but the final result for the axion-generated additional critical current $I_c^a$
as given by eq.~(\ref{ic1}) is actually independent of the precise value of this
formal $\vec{B}$-field, as it drops out of the equations.
Overall the effect of the galactic axion background is small but measurable and
the decaying axions
produce a small Shapiro step-like feature
when the axion mass resonates with the Josephson frequency.

We derived concrete formulas for the
additional critical current $I_c^a$ generated by axion flow, and derived conditions
for different types of Josephson junctions to act as optimum axion detectors. The measured voltage where
axion-generated Shapiro steps occur can be used to estimate the axion mass $m_a$, and their intensity
can be used to estimate the axionic dark matter density $\rho_a$ near the earth \cite{prl2013}.
We discussed peculiarities in the Shapiro step
patterns measured by four different experimental groups for very different types of Josephson junctions.
All four experiments point towards an axion mass of 110 $\mu$eV.
Further systematic measurements should still be performed  to test whether these candidate
signals are really due to axions. 
The axion mass value
of 0.11meV to which the
various Josephson experiments point to is compatible with current astrophysical and cosmological bounds on the axion mass.

\subsection*{Acknowledgement}
C.B. would like to thank Dr. Jason Robinson (Cambridge) for pointing out
Ref. \cite{golikova} to him.

\end{document}